# MAGNETOTRANSPORT IN TWO-DIMENSIONAL ELECTRON GASES ON CYLINDRICAL SURFACES

K.-J. FRIEDLAND, R. HEY, H. KOSTIAL, U. JAHN, E. WIEBICKE, K. H. PLOOG

*Paul-Drude-Institute for Solid State Electronics, Hausvogteiplatz 5-7, 10117 Berlin, Germany,*

A. VOROB'EV*[1], JU. YUKECHEVA[1], AND V. PRINZ[1]

*Institute of Semiconductor Physics, Russian Academy of Science, Acad. Lavrentyev Ave. 13, 630090 Novosibirsk, Russia*

We have fabricated high-mobility, two-dimensional electron gases in a GaAs quantum well on cylindrical surfaces, which allows to investigate the magnetotransport behavior under varying magnetic fields along the current path. A strong asymmetry in the quantum Hall effect appears for measurements on both sides of the conductive path. We determined the strain at the position of the quantum well. We observe ballistic transport in 8 μm wide collimating structures.

**1 Introduction**

Free-standing strained semiconductors are of considerable interest for new fundamental research fields, such as current-induced spin polarization [1], micromechanics with semiconductor cantilevers [2] and optics in microtube ring resonators [3]. One important new experimental approach is the self-rolling of thin pseudomorphically strained semiconductor bilayer systems as proposed by Prinz and coworkers [4]. This method allows the fabrication of free-standing lamellas or tubes of cylindrical shape with a constant radius of curvature based on epitaxial heterostructures grown by molecular beam epitaxy (MBE).

A promising application is the realization of heterostructures containing a two-dimensional electron gas (2DEG). Recently, tubes with a laterally structured 2DEG, e.g. in the form of a conventional Hall-bar with Ohmic and Schottky contacts on the unrolled part of the 2DEG were successfully fabricated [5,6]. This will allow to realize various structures with low-dimensional transport behavior on evenly curved surfaces. In the electrons' coordinate system, this is translated in a motion of the electrons along different and/or varying magnetic field lines with a given spatial dependence along the rolling direction of the semiconductor heterostructure according to the angle φ between the surface normal and the magnetic field $B_\perp = B_0 \cos(\varphi)$. An important parameter here is the low-temperature mean free path of the electrons $l_S = v_F \tau$, where $v_F$ is the Fermi

---

*A.V. acknowledges support by INTAS (Grant No. 04-83-2575) and the kind hospitality of the Paul-Drude-Institute

[1] Work partially supported by grant 04-02-16910 of the Russian Foundation for Basic research

velocity and $\tau$ the scattering time. Ballistic transport occurs, when $l_S$ is comparable to a typical system size $L$. In addition, to observe new physical phenomena which are related to the special motion of electrons, or motion of electrons in varying magnetic fields, $l_S$ should be comparable or even larger than the radius of the tube $r$, on which the electron is moving. For high mobilities heterostructures based on (Al,Ga)As systems are usually used. However, a significant draw-back for free-standing heterostructures on the basis of (Al,Ga)As systems with the formation of an inherent new surface is the Fermi-level pinning in the middle of the gap. This results in a significant depletion of the 2DEG, which is simultaneously accompanied by enhanced fluctuations of the potential and, therefore, by a reduced mobility in the 2DEG. Here we are faced with an obstacle which could not be circumvented so far. On the one hand, we need a large distance of the new surface from the 2DEG in order to avoid degradation of the electron mobility due to surface states. On the other hand, the heterostructure packets should be as thin as possible to realize a minimized rolling radius of the tube. Both requirements are contrary to each other, therefore ballistic transport on curved surfaces was not yet observed.

## 2  GaAs based high mobility 2DEG's on cylindrical surfaces

### 2.1  Screened high mobility 2DEG

We use a particular heterostructure, which was proven to significantly reduce the influence of the second surface on the mobility of the 2DEG in hybrid structures fabricated by epitaxial-lift-off and wafer-bonding [7].

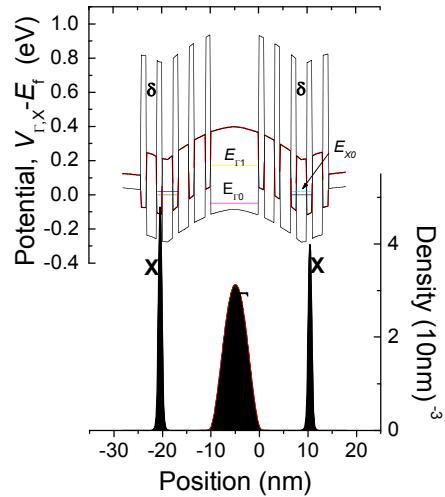

Figure 1. Potential and charge distribution in a GaAs SQW cladded by GaAs/AlAs SPSLs, calculated from a self-consistent solution of the Poisson and Schrödinger equations, including $\Gamma$ and X conduction band states. $\delta$ marks the positions of the $\delta$-doped layers.

The high-mobility 2DEG (HM2DEG) is located in a 13-nm-wide single quantum well (QW) with barriers consisting of short period AlAs/GaAs superlattices (SPSL), grown by (MBE) [8]. The superlattice period is chosen sufficiently short in order to ensure, that the X-like conduction-band states are lowest in energy in the AlAs compound of the SPSL. At sufficiently high doping concentration, these states are occupied with heavy mass X electrons, which are located close to the doping layer (Figure 1). The heavy mass of the carriers provides a high screening capability. Additionally, their Bohr-radius $a_B$ as well as their nominal distance from the doping layer is smaller or nearly equal to the average distance between the Si-atoms in the δ-like doping plane. Therefore, the X electrons can be very easily localized at the minima of the fluctuating potential, thus smoothing the fluctuations of the scattering potential (FSP). As a result, the mobility of the electrons in the GaAs SQW with X electrons in the SPSL can be considerably increased. The new concept for rolled-up films is that these X electrons also screen the high-mobility 2DEG from new surface states and its related FSP.

The HM2DEG package with an overall thickness of 192 nm was pseudomorphically grown on top of a 20 nm thick $In_{0.15}Ga_{0.85}As$ stressor layer forming the strained multi-layered films (SMLF). An additional 50-nm-thick AlAs sacrificial layer is introduced below the SMLF in order to release the SMLF from the substrate. Conventional Hall-bar structures with current flow along the [1 0 0] direction were fabricated photolithographically on the original substrate prior to the detachment process of the SMLF by shallow mesa etching and alloying of the Ohmic contacts. The magnetotransport measurements for a flat structure at a temperature of 0.3 K reveal a parallel conductivity in the GaAs and (In,Ga)As quantum wells. For the GaAs QW we estimate an electron density of $n_{GaAs} = 7.1 \times 10^{15}$ m$^{-2}$ with a mobility of about 120 and 50 m$^2$/Vs along the [1 $\bar{1}$ 0] and [1 0 0] directions respectively. The anisotropy of the mobility results from an anisotropic surface corrugation on the substrate [9].

### 2.2 Rolled-up 2DEG with contacts for magnetotransport measurements

Part of the SMLF containing the Hall-bar was detached by selective etching with an HF acid/water solution and directionally rolled up along the [100] direction in a tube with a visible outer radius of about 20-25 μm. Details of the fabrication process of the tubes are described in [10]. As a result, we obtain high-mobility 2DEG Hall-bar devices on rolled tubes. At low temperatures, we observe only one conductive channel, namely the one in the GaAs QW. The electron densities and mobilities range from $(5 - 7) \times 10^{15}$ m$^{-2}$ and (18 - 57) m$^2$/Vs respectively. The (In,Ga)As QW is fully depleted due to the close distance to the newly created surface.

#### 2.2.1 Ballistic transport in free-standing high-mobility 2DEG

For a device with the longest $l_S$, we chose transversal Hall-terminals for a nonlocal resistance measurement, when the electrons move ballistically, which resembles electron-beam collimation at two opposite quantum point contacts [11]. The rolled tube is oriented in such a way that the magnetic field is parallel to the surface normal at the Hall-cross position. The layout is schematically shown as an inset in Figure 2. Terminal 1 and 3 are source and collector contacts, respectively. A current of 50 nA is driven along the terminals 1 and 2, the voltage is measured between the terminals 4 and 3. The distance between source and collector is 8 μm, while the width of terminals 1 and 3 is 2 μm.

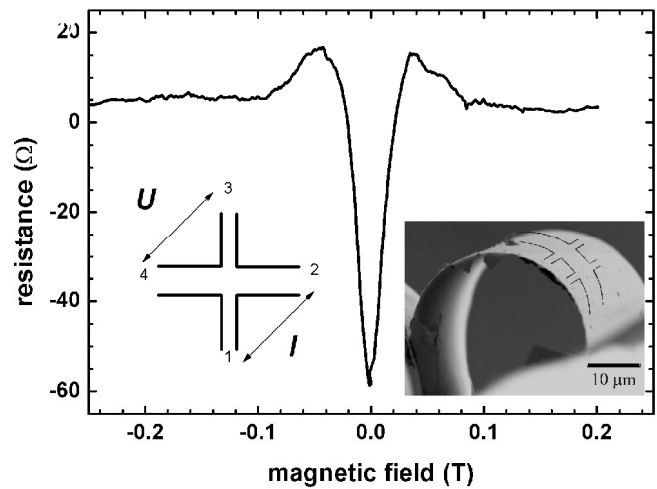

Figure 2. Bend resistance as a function of magnetic field at T = 0.3 K. The left inset displays the experimental layout. Right inset: Cross-sectional scanning electron micrograph of a rolled-up film. The position of the Hall-cross is schematically shown.

Due to the long $l_S$, electrons are collected at the opposite terminal at zero magnetic fields, causing a negative bend resistance (NBR). A small magnetic field bend off the 'beam' from the opposite terminal, thereby reducing the NBR [12]. At moderate magnetic fields, rebound trajectories of ballistic electrons cause a positive resistance, which disappears when all electrons are bent into the current terminal by high fields. This experiment demonstrates only the high quality of the 2DEG rather then carrying information about the ballistic transport in curved 2DEG structures. The ratio of the length of the mean free path to the rolling radius R is still rather low with a value less than 0.2.

2.2.2 Quantum Hall effect in high-mobility 2DEG's with varying magnetic field along the current direction

In a 16-μm-wide Hall-bar structure, we do not observe ballistic transport behaviour anymore. By orienting the Hall-bar with a current along the rolling direction, the measurement of the Hall resistances $R_{H1}$ and $R_{H2}$ at two adjacent transversal terminal pairs allow to derive the rolling radius $r$ through estimates of the local angles $\varphi_{Hi}$ between the field and the surface normals from $R_{Hi} = R_0 \cos(\varphi_{Hi})$. $R_0$ is the Hall resistance at a perpendicular magnetic field. If the magnetic field is perpendicular to the surface at a position within the adjacent transversal terminal pairs, the rolling radius $r$ can be determined from $\varphi_{H1} - \varphi_{H2} = l/r$, where $l$ is the distance between the Hall-terminals. For our tubes, we estimate $r$ to be about 18 μm. The rolling radius $r_{Th}$ can also be predicted from continuum strain theory using [13]

$$r_{Th} = \frac{h_1^4 + 4\chi h_1^3 h_2 + 6\chi h_1^2 h_2^2 + 4\chi h_1 h_2^3 + \chi^2 h_2^4}{6\varepsilon\chi(1+\upsilon)h_1 h_2 (h_1 + h_2)} \quad (1)$$

For the present layer system with the thicknesses of the stressor layer and the stack $h_1 = 18.7$ nm and $h_2 = 156$ nm, respectively, the ratio of Young's module $\chi$, the Poisson ratio $\upsilon$ and the strain $\varepsilon$, we calculate $r_{Th} = 23.9$ μm, which is close to, but systematically larger than the experimentally observed rolling radius $r$. From strain theory, we can estimate the strain at the position of the GaAs QW, which is compressive with a value $\varepsilon = 5 \cdot 10^{-4}$. At the original surface, the compressive strain is about one order of magnitude larger.

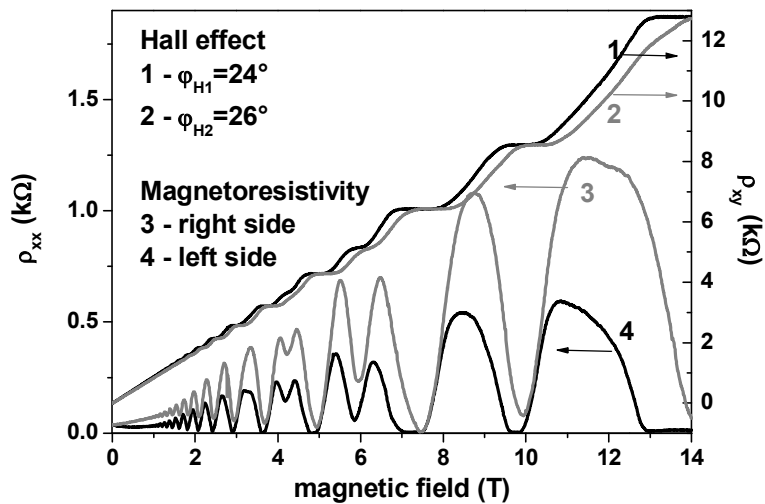

Figure 3. Transversal (1,2) and longitudinal (3,4) magnetoresistivities for the right and left sides of the Hall-bar on a cylindrical surface. The current and magnetic field directions are described in the text. T = 0.3 K.

Despite the rather large strain and the varying magnetic field projection over the current path, spin-split integer quantum Hall states remain observable for a perpendicular magnetic field at a position near the middle of adjacent transversal terminals, see Figure 3. Even the fractional Hall effect starts to develop at fields as high as 12 T. In this case, the difference between the Hall voltages remains sufficiently low and does not deform the equipotential lines too much across the conductor width. Nevertheless, even for this case, a strong asymmetry occurs in the magnetoresistivy for measurements on both sides of the Hall-bar, which we refer to as the right and the left sides in Figure 3. This asymmetry was interpreted in terms of current path bunching towards one of the Hall-bar edge [10] at classical magnetic fields, while for quantizing nonuniform magnetic fields an explanation was given [14], which takes into account the bending of the Landau edge states into the bulk.